\documentstyle[floats,twocolumn,aps,prl]{revtex}
\begin{document}
\ifpreprintsty\else
\twocolumn[\hsize\textwidth%
\columnwidth\hsize\csname@twocolumnfalse\endcsname
\fi
\title{Temperature dependence of spin polarizations at higher 
Landau Levels}
\author{Tapash Chakraborty$^\star$\nocite{byline}}
\address{Max-Planck-Institute for Physics of Complex Systems, 
D-01187 Dresden, Germany}
\author{Pekka Pietil\"ainen}
\address{Theoretical Physics, University of Oulu,
FIN-90570 Oulu, Finland}
\date{\today}
\maketitle
\begin{abstract}
We report our results on temperature dependence of spin
polarizations at $\nu=1$ in the lowest as well as in the next 
higher Landau level that compare well with recent experimental 
results. At $\nu=3$, except having a much smaller magnitude the 
behavior of spin polarization is not much influenced by higher 
Landau levels. In sharp contrast, for filling factor $\nu=\frac83$ 
we predict that unlike the case of $\nu=\frac23$ the system remains 
fully spin polarized even at vanishingly small Zeeman energies.
\end{abstract}
\ifpreprintsty\clearpage\else\vskip1pc]\fi
%\pacs{73.40.Hm,71.45.Gm,73.20,Dx}
\narrowtext

It has been long established that spin degree of freedom 
plays a very important role in the quantum Hall effects 
\cite{halloffame,book}, that are unique demonstrations of electron 
correlations in nature. At the Landau level filling factor 
$\nu=1$ ($\nu=n_e/n_s$, where $n_e$ is the electron number 
and $n_s=AeB/\hbar c=A/2\pi\ell_0^2$ is the Landau level 
degeneracy, $A$ is the area of the system and $\ell_0$ is 
magnetic length) the ground state is fully polarized with total
spin $S=n_e/2$ \cite{zhang}. A fully spin polarized state is also
expected for $\nu=\frac13$, while a spin unpolarized state is
predicted for the filling factor $\nu=2/m$, where $m$ is an odd
integer \cite{book}. Recently, a new dimension to those studies
was introduced by Barrett et al. \cite{sean} (see also \cite{manfra}) 
in their work on spin excitations around $\nu=1$ and also temperature
dependence of spin polarizations at $\nu=1$. Since then several
experimental groups have explored spin polarization of various other
filling factors. In these experiments, direct information about
electron spin polarization at various filling factors can be obtained
via nuclear magnetic resonance (NMR) spectroscopy. Information about
spin polarization of the two-dimensional electron gas in an externally
applied high magnetic field is derived here from measurements of Knight
shift of $^{71}$Ga NMR signal due to conduction electrons in GaAs quantum
wells in the quantum Hall effect regime. For a fully polarized ground
state, as is the case for $\nu=1$ and $\nu=\frac13$, experimental results
indicate that spin polarization saturates to its maximum value at very
low temperatures and drops rapidly as the temperature is raised
(Fig.~\ref{song1}, and also reported earlier in Ref.~\cite{sean,manfra}).
At large $T$, spin polarization is expected to decay as $T^{-1}$
\cite{tapash1,tapash2,tapash3} and was found experimentally to behave
that way \cite{sean,manfra}.

Recently, Song et al. \cite{song} reported NMR spectroscopy in a 
somewhat similar set up as that of Barrett et al. \cite{sean} in 
order to explore $\nu=1$ and $\nu=3$. Interestingly, temperature 
dependence of spin polarization at $\nu=3$ revealed a different behavior 
as compared to that at $\nu=1$. More specifically, the results of Song 
et al. indicated that even at the lowest temperature studied, electron
spin polarization at $\nu=3$ does not show any indication of saturation 
and with increasing temperature it sharply drops down to zero 
(Fig.~\ref{song2}). In this paper, we investigate spin 
polarization versus temperature at $\nu=1$ in the lowest Landau level 
as well as in the next Landau level. We also compare our results with 
experimental results of Ref.~\cite{song}. At low temperatures, the
behavior of spin polarization at $\nu=3$ is similar to that at
$\nu=1$ but of much smaller magnitude. These results agree reasonably
well with available experimental data at $\nu=3$. However, 
discrepancies between our theoretical results and the experimental 
data remain at higher temperatures. We also present theoretical results 
for $\nu=\frac23$ in the next higher Landau level. 
At $\nu=\frac23$, convincing evidence exist about the spin polarization 
in the lowest Landau level \cite{book,kukushkin,kukushnew}. 
But there are no experimental data available as yet for spin polarizations 
in the next higher Landau level, i.e., at $\nu=\frac83$. We find (somewhat
unexpectedly) that for $\nu=\frac83$, even at a vanishingly small 
Zeeman energy, electrons in the higher Landau level remain fully 
spin polarized.

We have calculated temperature dependence of spin polarization for 
different filling factors from \cite{tapash1,tapash2,tapash3},
$$\langle S_z(T)\rangle\equiv \frac1Z\sum\,{\rm e}^{-\varepsilon_j/kT}
\langle j|S_z|j\rangle$$
where $Z=\sum_j\,{\rm e}^{-\varepsilon_j/kT}$ is the canonical partition
function and the summation is over all states including all possible
polarizations. Here $\varepsilon_j$ is the energy of the state $|j\rangle$
with Zeeman coupling included. They are evaluated for finite-size
systems in a periodic rectangular geometry \cite{book}. Our earlier
theoretical results indicated that at small values of the Zeeman
energy, temperature dependence of spin polarization is non-monotonic 
for filling factors $\nu=2/m$, $m>1$ being an odd integer. In
particular, for $\nu=\frac23$ and $\nu=\frac25$, we found that
spin polarization initially {\it increases} with temperatures, reaching
a peak at $T\sim0.01 K$ when it falls as $1/T$ with increasing
temperature. Appearence of the peak was associated with spin transitions
at these filling factors and was found to be in good agreement 
with the experimental observation \cite{kukushkin}. For $\nu=1$
and $\nu=1/3$, our results are also in excellent agreement \cite{tapash2}
with the earlier available experimental results \cite{sean,manfra}.

\begin{figure}
\begin{center}
\begin{picture}(100,130)
\put(0,0){\includegraphics{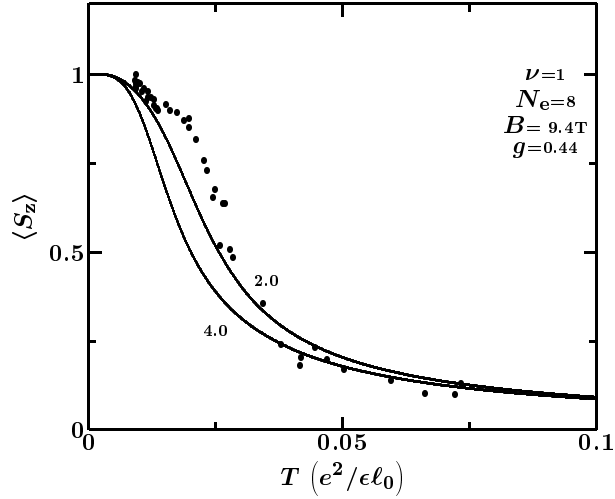}}
\end{picture}
\vskip 3.0 true cm
  \caption{Temperature dependence of spin polarization at $\nu=1$
  for two different values of finite-thickness parameter $\beta$. 
  Experimental data points are from Ref.~\protect\cite{song}. 
  }
  \label{song1}
\end{center}
\end{figure}

In our present work, energies are evaluated via exact diagonalization 
of a few electron system in a periodic rectangular geometry \cite{book}. 
Since even at the lowest experimental magnetic field the Landau level
separation $\hbar\omega_c$ is still an order of magnitude greater
than typical energies due to the Coulomb interaction, electrons in the 
lowest Landau level can be treated as inert. In the calculations that 
follow we can therefore consider the lowest Landau level to be an 
uniform background causing merely a constant shift to interaction energies. 
The higher Landau levels then enter the system Hamiltonian via a 
modified interaction potential \cite{highlandau}. More specifically, 
for a finite number of active electrons $N_e$ in a rectangular cell and 
choosing the Landau-gauge vector potential,
the Hamiltonian in the $n=0,1$ Landau levels is (ignoring the
kinetic energy and single-particle terms in the potential energy
which are constants \cite{book}),
\begin{eqnarray}
\nonumber
&&{\cal H}=\sum_{j_1,j_2,j_3,j_4}{\cal A}_{nj_1,nj_2,nj_3,
nj_4}a^\dagger_{nj_1}a^\dagger_{nj_2}a_{nj_3}a_{nj_4},\\
\nonumber
&&{\cal A}_{nj_1,nj_2,nj_3,nj_4}=\delta'_{j_1+j_2,j_3+j_4}
{\cal F}_n(j_1-j_4,j_2-j_3),\\
\nonumber
&&{\cal F}_n(j_a,j_b) = \frac1{2ab}\sum^{'}_{\bf q}\sum_{k_1}\sum_{k_2}
\delta_{q_x,2\pi k_1/a}\delta_{q_y,2\pi k_2/b}\delta'_{j_ak_2} \\
\nonumber
&&\times\frac{2\pi e^2}{\epsilon q}\left[
\frac{8+9(q/b')+3(q/b')^2}{8(1+q/b')^3}\right]\\
\nonumber
&&\times{\cal B}_n(q)\exp\left(\frac12q^2\ell_0^2-2\pi{\rm
i}k_1j_b/n_s\right),\\
\nonumber
&&{\cal B}_{n}(q) = \left\{ \begin{array}{ll}
			   1 & \mbox{for $n=0$,} \\
			   (1-\frac12q^2\ell_0^2)^2 & \mbox{for $n=1$,}
                      \end{array}
                 \right. \\
&&n_e=\left\{\begin{array}{ll}
		N_e & \mbox{for $n=0$,} \\
		\frac{\nu}{\nu-2}\,N_e & \mbox{for $n=1$.}
		\end{array} \right.
\nonumber
\end{eqnarray}
Here $a$ and $b$ are the two sides of the rectangular cell that contains
the electrons. The Fang-Howard variational parameter $b'$ is associated 
with the finite-thickness correction \cite{book}, $\epsilon$ is the 
background dielectric constant, and the results are presented 
in terms of the dimensionless thickness parameter
$\beta=(b'\ell_0)^{-1}$. The Kronecker $\delta$ with prime means that
the equation is defined ${\rm mod}\ n_s$, and the summation over $q$
excludes $q_x=q_y=0$. This numerical method has been widely used in the 
quantum Hall effect literature \cite{book} and is known to be very accurate 
in determining the ground state and low-lying excitations in the system.

\begin{figure}
\begin{center}
\begin{picture}(100,130)
\put(0,0){\includegraphics{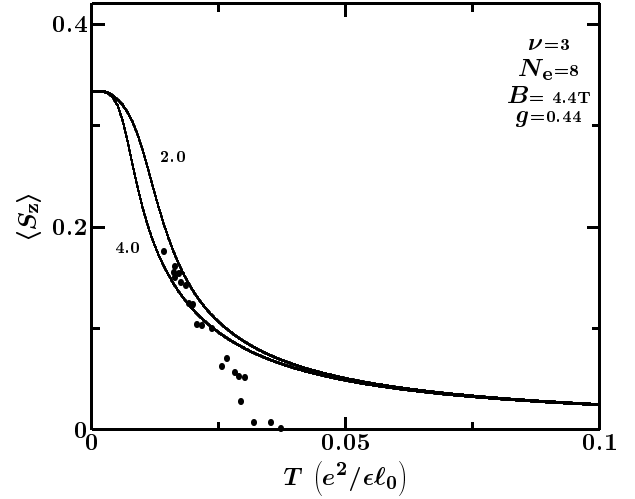}}
\end{picture}
\vskip 2.5 true cm
  \caption{Temperature dependence of spin polarization at $\nu=3$
  for different values of $\beta$. Experimental results are from
  Ref.~\protect\cite{song}. 
  }
  \label{song2}
\end{center}
\end{figure}

Our results for $\langle S_z(T)\rangle/{\rm max} \langle S_z(T)\rangle$
vs $T$ for an eight-electron system in a periodic rectangular geometry
at $\nu=1$ are presented in Fig.~\ref{song1} where we also
present the experimental data of Ref.~\cite{song} for comparison.
Here the temperature is expressed in units of $e^2/\epsilon\ell_0$ 
and the conversion factor to $K$ is $e^2/\epsilon\ell_0[K] =
51.67(B[{\rm tesla}])^{\frac12}$ appropriate for system studied
in experiments. In our calculations, we fix the parameters 
as in the experimental systems: the Land\'e $g$-factor is 0.44 and the
magnetic field is $B=9.4$ tesla. The curves that are close to the 
experimental data (and presented here) are for $\beta=2-4$. As we
discussed above, at low temperatures there is a rapid drop in spin
polarization and for high temperatures spin polarizations decay
as $1/T$. Our results are in good agreement with those experimental
features. They were also in good qualitative agreement with the earlier
experimental results at this filling factor \cite{tapash2}. While not
entirely new, these results are presented with the intention of comparing
them with the temperature dependence of spin polarization at $\nu=3$.
The results in the latter case are shown in Fig.~\ref{song2} (again for
an eight-electron system in a periodic rectangular geometry). In drawing
this figure, we have taken the following facts from the experimental
results of Ref.~\cite{song} into consideration:
(a) that the maximum $\langle S_z\rangle$ is in fact, 1/3 and not 
1 as in $\nu=1$, (b) the experimental scale at $\nu=3$ of Ref.~\cite{song}
is the same as that at $\nu=1$, and (c) spin polarization at $\nu=3$ 
is drawn in Fig.~\ref{song2} in the same scale as for $\nu=1$. All
the parameters except the magnetic field are kept the same as in the case
of $\nu=1$. Just as in the experimental situation, we fix the magnetic field
for $\nu=3$ at a much lower magnetic field of $B=4.4$ tesla. The filled
Landau levels, however, are still found to be inert at this low field
and does not influence our chosen Hamiltonian. As seen in
Fig.~\ref{song1}, numerical values of spin polarization are much
smaller here than those for $\nu=1$. Our theoretical results for
$\beta=2-4$ agree reasonably well with the experimental results of
Ref.~\cite{song} except in the high temperature regime where the
experimental data drop down to zero. Theoretical results, in contrast,
have the usual $1/T$ tail. We should point out however, that due 
to discreteness of the energy spectrum for finite number of electrons 
the terms with $S_z$ and $-S_z$ in the polarization cancel each other 
at high temperatures like $1/T$ and we will always end up with $1/T$ 
decay of $\langle S_z(T)\rangle$ vs $T$ \cite{tapash1}. 
Therefore, we cannot predict with certainty how a 
macroscopic system would behave at 
high $T$. However, given the fluctuations in data points
for $\nu=1$ and $\nu=3$ and the fact that the last few data points
for $\nu=3$ are extremely small, it is not clear if one expects
saturation of points with $1/T$ behavior or the spin polarization 
actually vanishes. Clearly, experimental data at high temperatures do not 
show any sign of saturation and in order to settle the question of actual
vanishing of $\langle S_z(T)\rangle$ it would be helpful to have more data 
in the high temperature regime. Saturation is also not visible in the 
low-temperature region of the experimental data. In order to clarify
many of these outstanding issues, it is rather important to
have more experimental probe of temperature dependence at this 
filling factor.

\begin{figure}
\begin{center}
\begin{picture}(100,130)
\put(0,0){\includegraphics{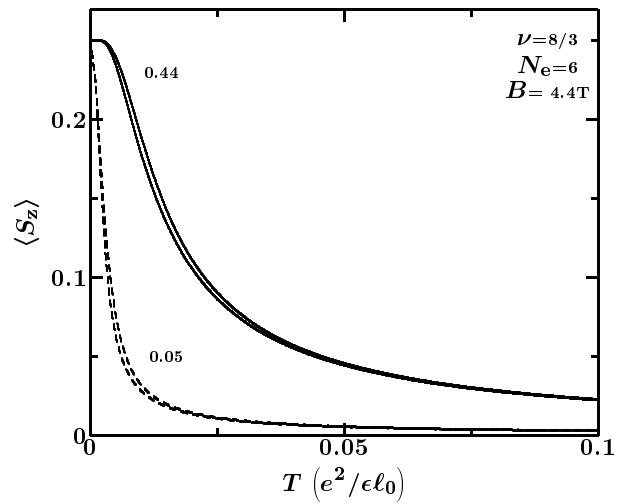}}
\end{picture}
\vspace*{2.5cm}
  \caption{Temperature dependence of spin polarization at $\nu=\frac23$
  for $\beta=2,4$ and two different values of Land\'e $g$-factor ($g=0.44$
  and $g=0.05$). The results are almost independent of $\beta$.
  }
  \label{spin}
\end{center}
\end{figure}

Influence of higher Landau levels is found to be quite significant for
filling factor $\nu=\frac23$. As we have demonstrated earlier
\cite{tapash1}, at low Zeeman energies the system at this filling 
factor is spin unpolarized and with increasing Zeeman energies, the
system undergoes a phase transition to a fully spin polarized state.
Similar result is also expected for $\nu=\frac25$. These theoretical
predictions are now well established through a variety of experiments 
\cite{kukushkin,kukushnew,engel,kron,kang,doroz}.
Our results for $\langle S_z(T)\rangle$ vs $T$ at $\nu=\frac83$ are shown 
in Fig.~\ref{spin}, where we present results for a six-electron system
and a magnetic field value of 4.4 tesla. In Fig.~\ref{spin}, we present 
our results for $\beta=2,4$, but the spin polarization is rather insensitive 
to the finite-thickness correction. We also consider two different values 
of Land\'e $g$-factor: 0.44 (solid curves) and 0.05 (dashed curves).
Interestingly, the results indicate that the total spin $S$ of
the active electrons, unlike in the lowest Landau level, is at its 
maximum value $S=N_e/2$ even without Zeeman coupling. Hence even an 
infinitesimal Zeeman coupling will orient the spins in the active 
system resulting the polarization to be $1/4$. That is at odds with the 
conventional composite fermion model which predicts fractions of the form
$2+2/m$, $m$ odd, to be unpolarized \cite{cfmodel}. This somewhat 
surprising behavior can be thought to be due to more repulsive effective 
interactions forcing the electrons, according to Hund's rule,
to occupy the maximum spin state more effectively as compared with 
electrons on the lowest Landau level. In order to demonstrate this 
behavior we have considered the case of a very small Zeeman energy 
(dashed curves), but the results still indicate full spin polarization 
of the active system. At this low Zeeman energy, spin
polarization drops rather rapidly from its maximum value as the
temperature is increased. In this context we should mention
that the idea of an extremely small Zeeman energy is not that
far fetched: in recent experiments, a significant reduction in
Zeeman energy has been achieved by application of a large
hydrostatic pressure on the heterostructure \cite{kang,press,holmes}.
It is even possible to have situations close to zero Zeeman energy
\cite{leadley}. With the help of all the different techniques available
in the literature to study spin polarization, it should be possible 
to explore $\langle S_z(T)\rangle$ for $\nu=\frac83$.

In closing, we have investigated spin polarization as a function of
temperature for $\nu=1$ and $\nu=\frac23$ in the higher Landau level.
Our results indicate that for $\nu=3$ our theoretical results are
not much influenced by the higher Landau level (except being much
lower in magnitude). Available experimental results are incomplete
at low and high temperature regions where no saturation of data points
have been observed. Our results at $\nu=\frac83$ reveal that the
system is always fully spin polarized even at very small Zeeman energies.
That is in contrast to the behavior at $\nu=\frac23$ which at low
Zeeman energies has a spin unpolarized state \cite{book} that is well 
supported by various experimental investigations. More experimental 
data points at $\nu=3$ in the low and high-temperature regime would be 
very helpful. Experimental probe of $\nu=\frac83$ with NMR and optical 
spectroscopy should be able to explore the spin states predicted in the 
present work.

We would like to thank Dr. Y.-Q. Song for sending us their 
experimental data and Igor Kukushkin for helpful discussions.

\end{document}